\documentclass{elsart4-1}

\usepackage{graphicx}

\usepackage{amssymb}

\usepackage[english,francais]{babel}

\newtheorem{e-proposition}[theorem]{Proposition}

\newtheorem{e-definition}[theorem]{Definition\rm}

\renewcommand{\theequation}{\arabic{equation}}
\newcommand{\zdep}{{z_{\tt dep}}} 
\newcommand{\zetadep}{{\zeta_{\tt dep}}} 
\newcommand{\zetaeq}{{\zeta_{\tt eq}}} 
\newcommand{\zetaff}{{\zeta_{\tt ff}}} 
\newcommand{\lopt}{{{\ell}_{\tt opt}}} 
\newcommand{\lrel}{{{\ell}_{\tt rel}}} 
\newcommand{\lav}{{{\ell}_{\tt av}}} 
\newcommand{\nudep}{{\nu_{\tt dep}}} 
\newcommand{\nueq}{{\nu_{\tt eq}}} 
\newcommand{\lrelax}{{{\ell}_{\tt relax}}}

\newcommand{\fig}[1]{Fig.~\ref{fig:#1}}

\def\og{\leavevmode\raise.3ex\hbox{$\scriptscriptstyle\langle\!\langle$~}}
\def\fg{\leavevmode\raise.3ex\hbox{~$\!\scriptscriptstyle\,\rangle\!\rangle$}}

\begin{document}

\centerline{Physics}
\begin{frontmatter}


\selectlanguage{english}
\title{Numerical Approaches on Driven Elastic Interfaces in Random Media}


\selectlanguage{english}
\author[authorlabel1]{E. E. Ferrero},
\ead{ferrero@cab.cnea.gov.ar}
\author[authorlabel1]{S. Bustingorry},
\ead{sbusting@cab.cnea.gov.ar}
\author[authorlabel1]{A. B. Kolton},
\ead{koltona@cab.cnea.gov.ar}
\author[authorlabel4]{A. Rosso}
\ead{alberto.rosso@u-psud.fr}

\address[authorlabel1]{CONICET, Centro Atomico Bariloche, 8400 S. C. de Bariloche, Argentina}
\address[authorlabel4]{Universite Paris-Sud, CNRS, LPTMS, UMR 8626, Orsay F-91405, France}


\medskip
\begin{center}
{\small Received *****; accepted after revision +++++}
\end{center}

\begin{abstract}
We discuss the universal dynamics of elastic interfaces in quenched random media.
We focus in the relation between the rough geometry and collective transport properties in
driven steady-states. 
Specially devised numerical algorithms allow us to analyze the equilibrium, creep, and depinning
regimes of motion in minimal models. 
The relevance of our results for understanding domain wall experiments is outlined.
{\it To cite this article:  E. E. Ferrero, S. Bustingorry, A. B. Kolton, A. Rosso, 
C.R. Physique XX (XXXX).}

\vskip 0.5\baselineskip

\selectlanguage{francais}
\noindent{\bf R\'esum\'e}
\vskip 0.5\baselineskip
\noindent
{\bf Approches Num\'eriques pour les interfaces \'elastiques en milieu al\'eatoire}
Nous discutons les principaux r\'esultats obtenus sur les propriet\'es universelles de la dynamique
des interfaces \'elastiques en milieu al\'eatoire.
Une attention particuli\`ere sera dedi\'ee \`a la relation entre la g\'eometrie rugueuse de l'interface
en mouvement et ses propiet\'es de transport collectif.
Les approches num\'eriques d\'evelopp\'ees permettent de d\'ecrire les propriet\'es d'\'equilibre, 
la dynamique de creep et la transition de d\'epi\'egeage de l'interface.
Nous discutons aussi la pertinence de nos r\'esultats dans les exp\'eriences sur la dynamique des parois.

{\it Pour citer cet article~: 
E. E. Ferrero, S. Bustingorry, A. B. Kolton, A. Rosso, C.R. Physique XX (XXXX).}

 \vskip 0.5\baselineskip
\noindent{\small{\it Mots-cl\'es~:} Reptation; D\'epi\'egeage; Desordre}}
\end{abstract}
\end{frontmatter}


\selectlanguage{english}
\section{Introduction}
\label{}
The out of equilibrium interplay between quenched disorder and elasticity in 
uniformly driven interfaces is at the root of the universal dynamical response 
displayed by very diverse physical systems.
Examples are 
magnetic~\cite{lemerle_domainwall_creep,bauer_deroughening_magnetic2,yamanouchi_creep_ferromagnetic_semiconductor2,metaxas_depinning_thermal_rounding,leekim_creep_corriente}
or ferroelectric domain walls~\cite{paruch_ferro_roughness_dipolar,paruch_ferro_quench,JoYang_creep_ferroelectrics,ferroelectric_zombie_paper,paruchbusting}, 
contact lines in wetting~\cite{moulinet_distribution_width_contact_line2,frg_exp_contactline}, 
fractures~\cite{bouchaud_crack_propagation2,bonamy_crackilng_fracture}, 
and earthquakes~\cite{jagla_kolton_earthquakes}. 
In this paper we discuss the basic phenomenology that emerges by solving, with specially devised numerical methods, 
the minimal models proposed for capturing such dynamical behavior.

Universal dynamical properties can be captured, both qualitatively and quantitatively, 
by rather simple models. To be concrete we will focus on the paradigmatic driven 
quenched-Edwards-Wilkinson (QEW) universality class, minimally described by \\
\begin{equation}\label{eq:eqmotion}
\gamma \partial_t u(x,t) = c \partial_x^2 u(x,t) + F_p(u,x) + f + \eta(x,t).
\end{equation}
This equation models the overdamped dynamics of an elastic interface with a univalued scalar 
displacement field $u(x,t)$, with $x$ a vector of dimension $d$, such that the interface is embedded 
in a space of dimension $D=d+1$. 
The elastic approximation made above, assumes small deformations such that the elastic part of the energy
is well described by $H_{el}=(c/2) \int_x dx^d [\nabla u]^2 $. 
We are thus ignoring ``plastic'' deformations such as overhangs or pinched-off loops that might
appear in real interfaces, and assuming that elastic interactions are short-ranged, with a 
stiffness constant $c$. 
The contact with a thermal bath at temperature $T$ is modeled by a Langevin noise, such that 
$\langle \eta(x,t) \eta(x',t') \rangle=2 k_B T\gamma \delta(x-x')\delta(t-t')$.
Finally, the interface is coupled to a uniform driving force $f$ and to a quenched pinning force $F_p$, which arises from the 
disorder in the host materials.
We will consider a non-biased pinning force characterized by its disorder-averaged correlator
\begin{equation}
\label{eq:correlator}
\overline{F_p(u,x)F_p(u',x')}=\Delta(u-u')\delta(x-x'),  
\end{equation}
with $\Delta(x)$ a short-ranged function, of range $r_f$ (determined 
by the domain wall width in real interfaces with point disorder).
For the so-called random bond (RB) case, the elastic line moves in a random short-range correlated potential.
The corresponding pinning force is $F_p(u,x) = -\partial_u U(u,x)$, where $U(u,x)$ is the random potential,
and thus $\int_u \Delta(u,x)=0$. 
For the random field (RF) case, $U(u,x)$ is a random walk 
as a function of $u$, with diffusion constant $\int_u \Delta(u,x)>0$~\cite{chauve_creep_long}.

The model just defined is minimal, and in order to compare with experiments other ingredients might be considered.
For example in charge density waves and vortices, the elastic structure is 
periodic~\cite{giamarchi_vortex_review,nattermann_cdw_review}, while in 
fracture~\cite{bonamy_crackilng_fracture,gaoandrice,tanguy_longrange,bonamy_review,alava_review_cracks} and 
wetting~\cite{degennes_joanny,ledoussal_contact_line} the elastic interactions are long ranged. 
Moreover, anharmonic corrections to the elasticity or anisotropies can be also 
relevant~\cite{rosso_dep_exponent,tang_anisotropic_depinning}. 
Remarkably, all these different universality classes (with different critical exponents) share the same basic physics
of the QEW model discussed in the following sections.

\section{Transport and Geometry of Driven Interfaces in Random Media}
\label{sec:transportandgeometry}
\begin{figure}[!tbp]
\centering
\includegraphics[width=0.95\textwidth]{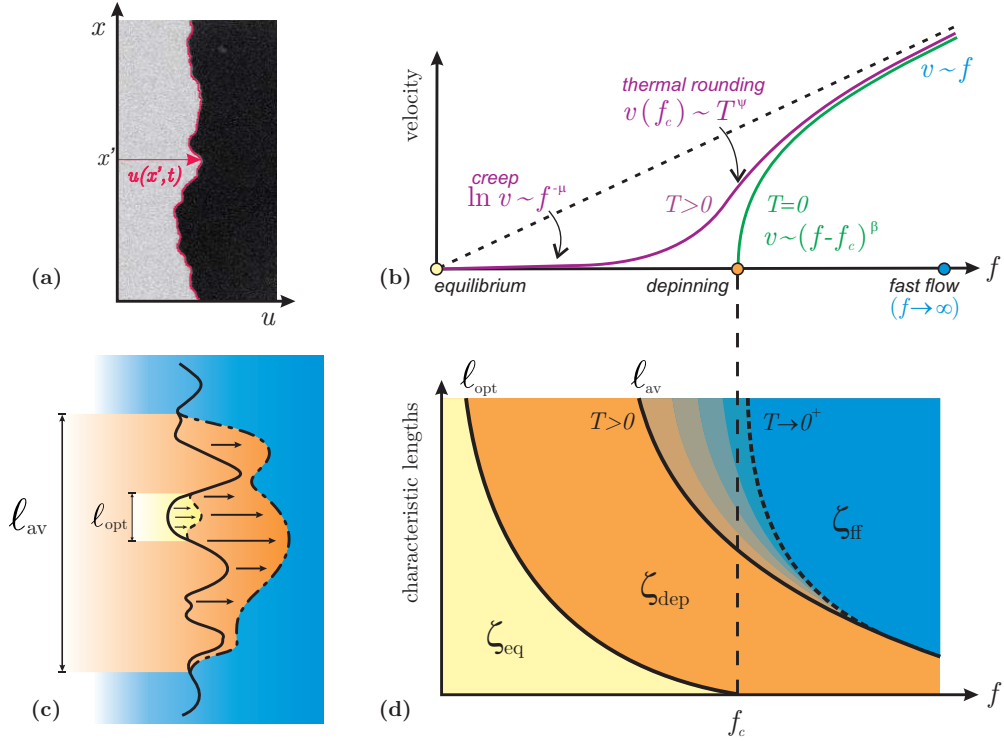}
\caption{\label{fig:whatcanwesay}
Linking transport and geometry.
(a) Snapshot of a domain wall in a two dimensional ferromagnet~\cite{cortesiaferre}.
(b) Typical velocity-force characteristics.
(c) Crossover lengths $\lopt$ and $\lav$ representing 
the optimal excitation and the deterministic avalanches, respectively.
(d) Geometric crossover diagram. 
}
\end{figure}

Disorder makes the interface dynamics very rich. 
On one side, the interface appears rough, both in absence and in presence of  drive. 
On the other side, the response of the system to a finite drive is strongly nonlinear 
and, at zero temperature, motion exists only above a finite threshold, the so-called 
critical force $f_c$.
As summarized in ~\fig{whatcanwesay} collective dynamics is observed in different regimes: just above $f_c$ the motion
is very jerky, and displays collective rearrangements or avalanches of a typical size 
$\lav$.
Below $f_c$ motion is possible only by thermal activation over energy barriers and 
the interface slowly fluctuates backward and forward.
This ``futile'' and reversible dynamics takes place up to a characteristic 
size\footnote{The subscript ${\tt opt}$ stays from ``optimal'' because $\lopt$ is the
size of the jump associated with the optimal barrier that the interface should overcome 
in order to find a new metastable state with a smaller energy.} 
$\lopt$ above which energy barriers do not exist before the next lower-energy metastable state and the interface moves only 
forward pulled by the finite drive $f$.
As $f \to 0$ both  $\lopt$ and the energy barriers diverge.

\subsection{Three reference stationary states}
\label{sec:referencestates}
Universality manifests, both in the transport and in the geometry of the moving interface, 
through the existence of robust critical exponents.
These describe the rate of power-law divergences of important quantities as the control 
parameter $f$ approaches three special states:
(i) the {\it equilibrium} ($f=0$); (ii) the {\it depinning} ($f=f_c$, $T=0$), 
and the {\it fast-flow} ($f \gg f_c$). 
Beyond some microscopic length-scale~\cite{agoritsas2010,agoritsas_review} 
these three steady-states have the peculiarity that 
the geometry is self-affine with their characteristic roughness exponent: 
(i) $\zetaeq$, (ii) $\zetadep$, and (iii) $\zetaff$.
Self-affinity means that length $x$ and displacement $u$ can be rescaled
as $x'=ax$ and $u'=a^{-\zeta} u$ into a new interface $u'(x')$ which is
statistically equivalent to $u(x)$, namely $u(a x) \sim a^{\zeta} u(x)$.
Transport properties in these three points are very different:
At {\it equilibrium}, the mean velocity is zero and the dynamics is glassy: at small length 
scales, we observe a fast futile motion, while in order to observe a rearrangement on a 
large scale size $l$ we need to overcome a barrier $E_b(l)$ growing as $E_b\sim l^\theta$,
with $\theta$ a positive and universal exponent.
Dimensional analysis suggests that $\theta=d-2+2\zetaeq$, which is confirmed by accurate 
numerical studies~\cite{drossel}.
At the zero temperature {\it depinning transition } the velocity vanishes as 
$v(f,T=0) \sim (f-f_c)^\beta$ for $f>f_c$ while $v=0$ for $f<f_c$.
At finite temperature, this sharp transition is rounded
and the velocity behaves as $v(f_c,T) \sim T^\psi$.
This is the so-called {\it thermal rounding regime}. 
At large force, $f\gg f_c$, in the {\it fast-flow regime}, we recover the linear response 
$v \sim f$.
Here impurities generate an effective thermal noise on the interface with 
$T_{\tt eff}-T \sim \Delta(0)/v$ (the disorder strength  $\Delta(0)$ is defined 
in Eq.(\ref{eq:correlator})).
Therefore, the fast-flow roughness corresponds to the Edwards-Wilkinson 
roughness\footnote{In real experiments we expect non-linear terms to become relevant and
change the Edwards-Wilkinson into the Kardar-Parisi-Zhang universality class} $\zetaff=(2-d)/2$.
The three reference state are schematically represented in \fig{whatcanwesay}(b).

\subsection{Connecting the three reference stationary states: creep and depinning}
\label{sec:creepanddepinning}
How does the interface behaves in between (i.e., $0<f<f_c$ and $f>f_c$) these three reference
steady-states?
Let us consider some instantaneous snapshots of an interface as obtained from an
experiment or a numerical simulation.
\fig{whatcanwesay}(a) is an illustrative experimental example for a ferromagnet~\cite{cortesiaferre}.
Let us assume that the longitudinal size of the snapshot is large enough to contain all the 
relevant characteristic length-scales, and that the driven interface is already in a stationary 
regime so that the memory of the initial condition is lost.
By analyzing the snapshot, what can we say about the interface state of motion?

One could imagine a naive scenario where the dynamic roughness exponent $\zeta$ varies continously
upon increasing the driving force, from its value at the equilibrium $\zetaeq$, to $\zetadep$,
and finally to $\zetaff$.
Actually, this is not true, and the interface geometry can be instead described by only these three
exponents and by the two crossover lengths $\lopt$ and $\lav$.
The crossover or dynamical phase diagram is schematically shown in \fig{whatcanwesay}(d).
At small temperatures and below threshold, $0<f<f_c$,  the interface is in the ultra-slow {\it creep regime}.
At small length-scales $\ell < \lopt$ the interface looks like an equilibrated interface 
(with an exponent $\zetaeq$).
For intermediate scales, $\lopt < \ell < \lav$, we expect the same roughness exponents as those of the depinning
transition, $\zetadep$.
Finally, at the largest length-scales, $\lav < \ell$, we expect to measure the fast-flow roughness
exponent $\zetaff$.
Let us observe that the large length scales are controlled by non-equilibrium roughness  exponents.
This shows that even for very small forces $f$, the system is {\it far from 
equilibrium}~\cite{chauve_creep_long,kolton_depinning_zerot2,kolton_dep_zeroT_long}, in contrast 
with the initial physical pictures that  assumed metastable configurations indistinguishable from
the equilibrium ones.
Analogously, for $f>f_c$ we observe that at short length-scales $\ell<\lav$ the roughness exponent 
is $\zetadep$, while for $\ell > \lav$ one has $\zetaff$.

The dynamical phase diagram of \fig{whatcanwesay}(d) thus allows us
to get important information from snapshots such as the one of \fig{whatcanwesay}(a):
(i) It can tell us first if the interface of the snapshot was in the $f<f_c$ creep, in the 
$f \gtrsim f_c$ depinning, or in the $f \gg f_c$ fast-flow regimes;
(ii) The actual values of the different roughness exponents (obtained by fitting, for instance,
the structure factor $S_q$) guide us in the search for a universality class;
(iii) Looking at snapshots for different forces could give access to the critical behavior of 
$\lopt$ and $\lav$, namely $\lopt\sim f^{-\nueq}$ and $\lav(T=0) \sim (f-f_c)^{-\nudep}$. 
At finite temperature, $\lav$ remains finite below threshold but quickly diverges as 
$f\to 0$~\cite{chauve_creep_long}.

Interestingly, both $\lav$ adn $\lopt$ have a ``double'' physical meaning: besides being roughness
crossover lengths, they control the non-linear collective transport properties of the interface.
Indeed, in the depinning regime $f \gtrsim f_c$, $T=0$, the jerky motion can be characterized by 
a velocity autocorrelation with finite spatial and time ranges $\lav$ and $\lav^{\zdep}$, 
respectively (with $\zdep$ the dynamical exponent).
Since the width of the avalanches is expected to grow as $\lav^{\zetadep}$ the velocity can be thus
estimated as $v \sim \lav^{\zetadep-\zdep}$.
On the other hand $v \sim (f-f_c)^\beta \sim \lav^{-\beta/\nudep}$, yielding the hyperscaling 
relation $\beta=\nudep(\zdep-\zetadep)$.

In the $f < f_c$ creep regime, three different dynamical steps can be isolated
(see Sec.\ref{sec:creepregime}):
(i) Starting from  a deep metastable state, the interface explores the neighborhood jumping back and
forth (futile dynamics);
(ii)  This continues until a saddle configuration  is found.
The saddle configuration will differ from the initial metastable state on a typical length $\lopt$;
(iii) Finally, from the saddle configuration the interface  relax {\em deterministically} to a new
and deeper metastable configuration.
The size of the optimal excitation $\lopt$ can be estimated by balancing the  gain in energy of being
pinned in a deep metastable state, $E_{{\tt metast}}(\ell)$, with the gain in energy of  moving the
interface forward  $\sim f \, \left( u_{\tt{saddle}}(\ell)-u_{\tt{metast}}(\ell) \right)$.
At equilibrium  we have that $u_{\tt{saddle}}(\ell)-u_{\tt{metast}}(\ell) \sim \ell^{\zetaeq}$ and 
that $E_{\tt{pinned}}(\ell) \sim \ell^\theta$, with $\theta=d-2+2 \zetaeq$.
The balance gives $\lopt \sim  f^{-\nueq}$, with $\nu_{eq} =1/(2-\zetaeq)$, which means that $\lopt$
diverges when  $f \to 0$, as expected.
Moreover, numerical simulations give a clear evidence that the energy landscape is characterized by
a unique energy scale, and  that the energy difference between neighbor metastable states is equal to
the energy barrier separating them \cite{drossel}.
Thus, we expect   the barrier that  the interface has to overcome to escape from a deep metastable
state to grow as  $\lopt^\theta$. Using the Arrhenius activation, we recover the creep formula:
\begin{equation}
\label{creep_formula}
v \sim \exp(- C\,  f^{-\mu_{\tt{eq}}}/T)
\end{equation}
where $\mu=(2 \zetaeq-1)/(2-\zetaeq)$ is an equilibrium exponent.
This formula has largely been verified by experiments on magnetic domain
walls~\cite{lemerle_domainwall_creep,metaxas_depinning_thermal_rounding,leekim_creep_corriente}.
When the barriers are too small compared with $T$, the velocity is no longer described by the Arrhenius
law, leading to a thermal rounding behaviour  $v \sim \lav^{-\beta/\nudep}$, with
$\lav \sim T^{-\psi \nudep/\beta}$ at $f=f_c$. For harmonic short range elasticity it can be shown that
$\nudep=1/(2-\zetadep)$ (Statistical Tilt Symmetry relation~\cite{fisherhuse_sts}).

As summarized in \fig{whatcanwesay}, geometry and transport are thus closely related.
In the following sections we explain how the above phenomenology can be numerically obtained.

\section{Method and Observables}

The main difficulty with Eq.~(\ref{eq:eqmotion}) is the non-linearity introduced by the disorder, 
which breaks the translational symmetry.
Indeed, in the absence of disorder Eq.~(\ref{eq:eqmotion}) becomes the Edwards-Wilkinson equation
which is exactly solvable~\cite{edwards_wilkinson,krug_review}. 
With disorder, the mean field model, valid above the upper critical dimension $d_{\tt uc}=4$ was
solved by Fisher \cite{fisher_depinning_meanfield}.
Advanced analytical techniques such as the functional renormalization group (FRG) allows to obtain
expansions below but close to $d_{ \tt uc}$~\cite{chauve_creep_long}.
Numerical approaches thus appear as a valuable and necessary theoretical tool.
From now on we will focus on the $d=1$ QEW model, which is:
(i) experimentally relevant for interfaces in thin films,
(ii) a stringiest case for testing analytical approaches in principle valid only close to $d_{\tt uc}$,
(iii) simpler to tackle numerically at large length scales.

Numerically, it is  convenient to  discretize the interface in the $x$-direction, keeping $u(x,t)$
as a continuum variable\footnote{For developing some algorithms and specially for the statics, it
might be also convenient to discretize the displacement field $u(x,t)$.}.
The center of mass velocity for an interface of size $L$ is defined as,
\begin{equation}
 v(t) = \overline{\frac{1}{L} \sum_{x=0}^{L-1} \partial_t u(x,t)},
\end{equation}
\noindent that, given Eq.~\ref{eq:eqmotion} and $\eta=c=1$, is nothing else but the spatial average of 
the instantaneous total forces acting on the line.
The geometrical properties of the line as a function of length-scale can be conveniently described 
using  the instantaneous quadratic  width:
\begin{equation}
 w^2(t) = \overline{\frac{1}{L} \sum_{x=0}^{L-1} [u(x,t)-u(t)]^2},
\end{equation}
\noindent or the averaged structure factor
\begin{equation}\label{eq:Sq}
S_q(t) = \overline{\biggl|\frac{1}{L}\sum_{x=0}^{L-1} u(x,t) e^{-iqx}\biggr|^2} ,
\end{equation}
\noindent where $q=2\pi n/L$, with $n=1,\ldots,L-1$, and $u(t)=L^{-1} \sum_{x=0}^{L-1} u(x,t)$ is
the center of mass position.
When the steady state regime is reached, these quantities become time independent.
In particular, for the three reference states (equilibrium, depinning and fast flow), the last two
quantities display self-affine behavior: $w^2 \sim L^{2 \zeta}$ and $S_q \sim 1/q^{d +2 \zeta}$.
For different values of the external force, the above observables display  crossovers between the 
corresponding reference states.

\section{Numerical methods and main results}
\label{sec:numericalmethods}

At zero force the steady state is the equilibrium state which corresponds to the ground state at $T=0$.
There exist exact and efficient numerical methods for targeting the equilibrium  both at
zero and finite temperature.
These methods rely on  transfer matrix techniques and apply to short range
elasticity~\cite{huse_henley_polymer_equilibrium}.
Upon increasing the driving force from zero, we enter first in the {\it creep regime} which is defined
at low, but non zero temperature.
In this regime the Langevin dynamics can be implemented at finite temperature \cite{chen_marchetti,kolton_creep2},
but becomes highly inefficient in the vanishing temperature limit ($T \to 0^+$).
Fortunately, in this limit an exact algorithm allows to characterize the ultra slow motion of the
line~\cite{kolton_depinning_zerot2,kolton_dep_zeroT_long}.
For a larger value of the external force we reach the depinning transition ($f=f_c$).
At $T=0$ direct Langevin dynamics simulations converge very slowly to the correct steady-state.
Different methods have been thus developed:
(i) an efficient algorithm  allows an exact  calculation of the critical force and critical configuration
for each disorder realization~\cite{rosso_depinning_longrange_elasticity}.
(ii) Molecular dynamics simulations of the {\it non-steady} relaxation allow to get the dynamical critical
exponents and  target  the thermodynamic critical force~\cite{FeBuKo2013}.
At $T>0$ and $f=f_c$, Langevin dynamics allows to describe the critical thermal rounding of the depinning
transition~\cite{bustingorry_thermal_rounding_epl,bustingorry_thermal_depinning_exponent}.
Finally simple Langevin dynamics in the large force limit ($f \gg f_c$) captures the features of the fast
flow regime at all temperatures.

\subsection{Equilibrium State}
\label{sec:equilibrium}

At zero force, the geometrical properties of the elastic string can be targeted using a transfer matrix
method for the discrete directed polymer model.
The line is described by the discrete variable $u(x)$, that gives the displacement of the string on the slice
$x$ of a square lattice, and a disorder potential $U$, which is drawn on each site $(x,u)$.
The energy of a given configuration is given by the sum of the site energies along the path $u(x)$: 
$E=\sum_x U(x, u(x))$.
Furthermore, the so-called solid-on-solid (SOS) restriction ${|u(x+1)-u(x)|=1}$ has to be implemented.
When $f=0$ this model belongs to the same universality class as Eq.~(\ref{eq:eqmotion}).

Given a disorder realization, the ground state configuration of a polymer starting in $(0,0)$ and ending
in $(x,u)$ can be found using the following recursive relation:
\begin{equation}
E_{\tt min}(x,u) = U(x,u) + \min
[  E_{\tt {min}}(x-1,u-1/2), E_{\tt {min}}(x-1,u+1/2)
].
\end{equation}
At finite temperature $T$ it is possible to compute the weight  $Z_{x,u}$ of all polymers starting in 
$(0,0)$ and ending in $(x,u)$  using the following recursion:
\begin{equation}
\label{eq:Zeta}
 Z_{x,u} = e^{-U(x,u)/T} \left( Z_{x-1,u-1/2}+ Z_{x-1,u+1/2}\right)
\end{equation}
with the initial condition $Z_{0,u} = \delta_{0,u}$.
Therefore, the probability to observe a polymer ending in $(x,u)$ is $Z_{x,u}/\sum_{u'} Z_{x,u'}$.
Due to the recursion relation, $Z_{x,u}$ grows exponentially with $x$. 
To avoid numerical instability, all weights $Z_{x,u}$ at fixed $x$ have to be divided by the largest one, 
which does not change the polymer ending probability \cite{bustingorry2010}.
With these procedures it is possible to characterize both the geometrical properties of the polymer
(as the roughness exponent $\zetaeq=2/3$ in 1d) and the free energy or ground state energy fluctuations
(as the characteristic exponent $\theta$)~\cite{prahofer2002,agoritsas_free_energy1}.

\subsection{Creep regime}
\label{sec:creepregime}

\begin{figure}
\centering
  \includegraphics[width=0.95\textwidth]{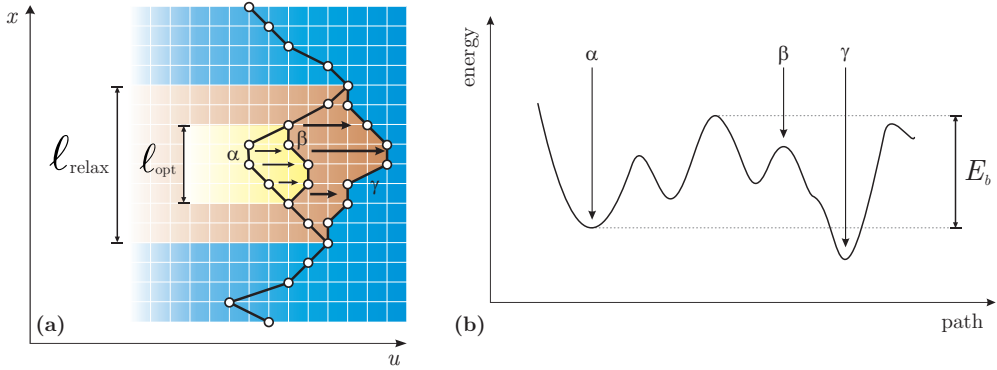}
  \caption{Low temperature dynamics of the driven elastic string below the depinning threshold. 
The optimal path to escape from a given metastable configuration $\alpha$ pass through a saddle
configuration $\beta$, from which the system is able to relax deterministically to the next metastable configuration 
$\gamma$ with $E_{\gamma} <E_{\alpha}$.
  }
\label{fig:creep}
\end{figure}

For non-zero drive but still below the depinning threshold ($0<f<f_c$), a finite temperature must be imposed 
in order to forget the initial condition and reach a unique stationary state. 
In references~\cite{kolton_depinning_zerot2,kolton_dep_zeroT_long} the regime relevant for {\it creep motion}
was studied; this is, the steady state regime of an interface in the limit of vanishing temperature.
A direct numerical study of the motion of such regime is very difficult.
The motion takes place by thermal activation over barriers leading to extremely 
long activation times making numerical techniques such as the Langevin dynamics inefficient.
Fortunately, different numerical method~\cite{kolton_depinning_zerot2,kolton_dep_zeroT_long} 
allows to follow the  motion of an interface at finite temperature, avoiding 
the above-mentioned difficulty.
This method directly implements the polymer dynamics as a {\it unique forward-moving sequence of metastable
states of decreasing energy}.
It can be proved that such a sequence corresponds to the exact dynamics for a finite interface in a given
disorder realization in the $T\to 0$ limit.

The elementary step of such a dynamics is sketched in \fig{creep}.
It consists in finding the {\it optimal path} from one metastable configuration $\alpha$ with energy
$E_{\alpha}$ to the next  metastable configuration $\gamma$, with a lower energy.
This path pass through a {\it saddle} configuration $\beta$ with an energy $E_{\beta}> E_{\alpha}$.
Once in  $\beta$ the polymer relaxes deterministically to $\gamma$.
The configurations $\alpha$ and $\gamma$ differ from each other by a portion of size $\lrelax$, while
$\alpha$ and $\beta$ differ by $\lopt$.
The maximun energy  $E_{\tt max}$ reached by the polymer through the path defines a drive dependent
barrier $E_b(f)=E_{\tt max}- E_{\alpha} $, therefore the time  associated to this elementary step is
given by $ \sim e^{E_b(f)/T}$.

\begin{figure}
\centering
  \includegraphics[scale=0.85]{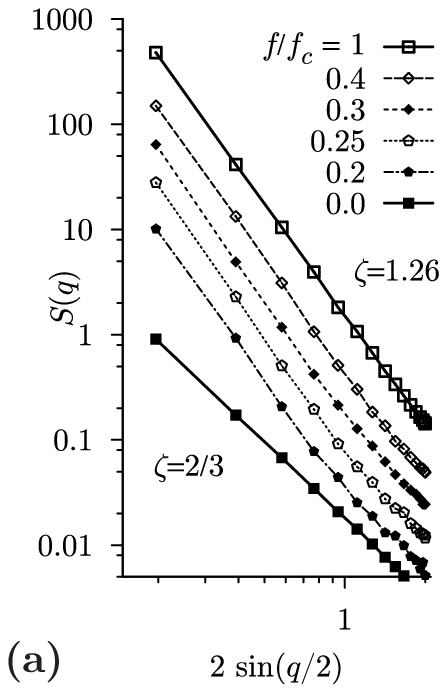} 
  \hfill 
  \includegraphics[scale=0.85]{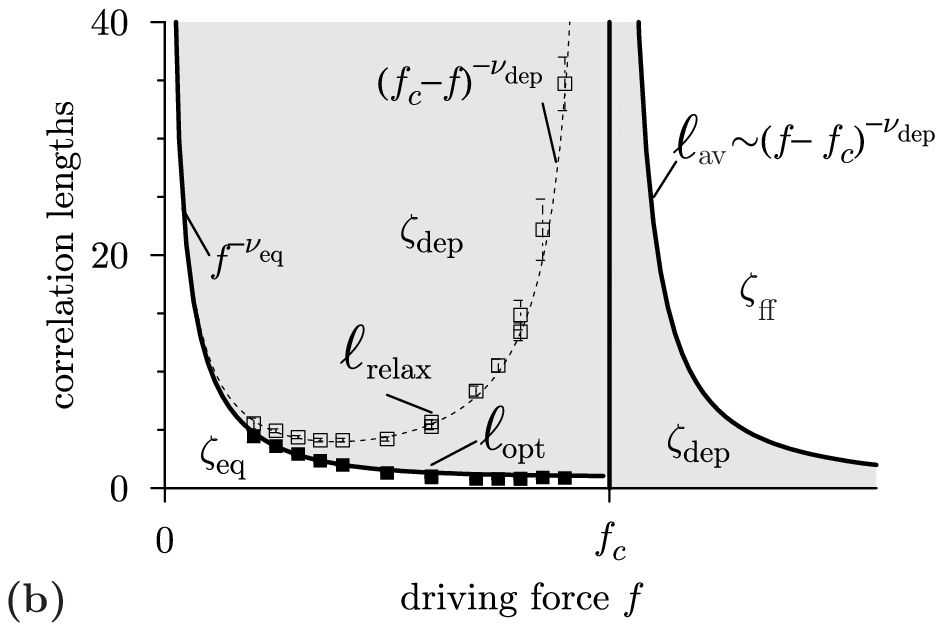}     
\caption{ (a) Steady state structure factor $S(q)$ of the line in the $T \rightarrow 0$ limit for 
different forces (curves are shifted for clarity).
(b) The steady state properties of the elastic string at $T\rightarrow 0$ are determined by 
$\lopt$ (filled symbols) and by  $\lav$.
These lengths separate regions characterized by the equilibrium exponent, the depinning exponent 
(gray region), and the  fast flow one. The divergent length $\lrelax$ (open symbols) is associated
only with transient dynamics. Lines are guides to the eye.
\textit{Figures are adapted with permission from}~\cite{kolton_depinning_zerot2}.
}
\label{fig:phasediagram}
\end{figure}

Numerical results are summarized in \fig{phasediagram}:
(i) The structure factor  behavior below $f_c$ supports the picture proposed in \fig{whatcanwesay}.
In particular this behavior is consistent  with FRG calculations~\cite{chauve_creep_long} that predict
in the creep regime  the existence  of a characteristic scale (namely $\lopt$) below which the system
is at equilibrium and above which the system is characterized by deterministic forward motion.
Moreover, in qualitative agreement with these predictions it is observed that both $\lopt$ and $E_b(f)$
increase as $f$ decreases.
Unfortunately this algorithm has not allowed to explore the region of vanishing $f$ where $\lopt$ is
expected to diverge as  $\sim f^{-1/(2-\zetaeq)}$ and the ``optimal'' barrier as $E_b(f) \sim f^{-\mu}$.
In turn, it can be seen that the size $\lrel$ diverges as $\sim (f_c-f)^{-\nudep}$ and this is 
interpreted as the linear size of the deterministic depinning-like avalanche triggered by  thermal
nucleation.
It is worth remarking that although $\lrelax$ diverges as $f \to f_c$  from below, it does not affect
the steady-state spatial correlations, unlike  $\lav$ and $\lopt$.

In order to study the effects of larger temperatures one can use Langevin dynamics \cite{kolton_creep2}.
These simulations show that $\lav$ has a finite value for $T>0$ and $0<f<f_c$ and it actually diverges 
as $f \to 0$ (as sketched in \fig{whatcanwesay}), but the lack of precison does not
allow to test the universal behavior $\lav \sim T^{-\sigma} f^{-\lambda}$ predicted by FRG
calculations~\cite{chauve_creep_long}. Note that $\lav$ diverges as the velocity goes to 
zero, while $\lrelax$ remains finite in the same limit. In other words, while in the $T\to 0$ 
limit the string is blocked in the first metastable state with lower energy 
producing an avalanche of typical size $\lrelax$, 
at finite temperature a larger avalanche of size $\lav$ takes place. In the creep regime 
these depinning-like processes are fast in comparison with the waiting times for 
the thermally activated jumps ($\lopt$).

Finally,  the finite temperature long time  relaxation of a flat line {\it in absence of drive} can be
described as a creep process over barriers that increase in time, such that the growing correlation
length (separating the equilibrated length-scales with the ones retaining memory of the initial condition)
is~\cite{kolton_flat_interface} $\ell(t) \sim [T \log(t)]^{1/\theta}$.

\subsection{Depinning regime at $T=0$}
\label{sec:depinning}

\begin{figure}[!tbp]
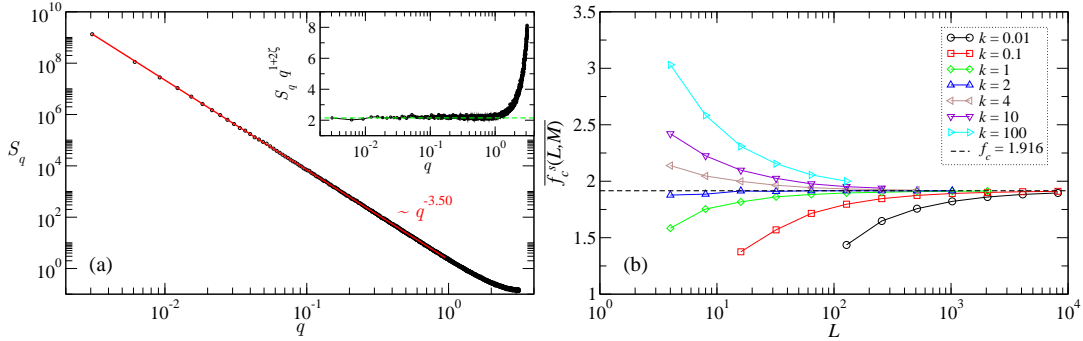

\includegraphics[scale=0.29,clip=true]{figures/criticalSq}
\hfill
\includegraphics[scale=0.29,clip=true]{figures/criticalFc}
\caption{\label{fig:criticalstate} 
Results for the critical state in finite samples obtained with the exact algorithm of 
Sec.~\ref{sec:criticalforce}.
(a) Critical configuration structure factor (see definition in Eq.~(\ref{eq:Sq})). 
The fitted roughness exponent yields $\zetadep=1.250\pm 0.005$.
(b) The finite-size critical force $f_c$ as a function of the longitudinal size $L$ for periodic samples 
of size $L\times M$ with $M=kL^\zetadep$.
As $L$ grows the asymptotic value of the critical threshold becomes independent on the aspect ratio $k$.
The dashed line corresponds to the thermodynamic limit $f_c = 1.916 \pm 0.001$.
\textit{Figures are reproduced with permission from}~\cite{FeBuKo2013}.
} 
\end{figure}

\subsubsection{Critical force and critical configuration}
\label{sec:criticalforce}

At $T=0$, the transition between a pinned phase and a moving phase is displayed in any finite sample
with a given disorder realization.
This is assured by a set of properties pointed out by Middleton \cite{middleton_theorem}.
The first one is  the so-called ``no-passing rule'':
if two strings $u(x,t)$ and $\tilde u(x,t)$ do not cross at a given time, they will not cross at any
later times. 
Another important property of Middleton's theorem states that if, at an initial time, the velocities 
are non-negative for all points $x$, they will remain non-negative for all later times.
It follows from these properties that, once we have found a forward moving string, we can be sure
that snapshots of the string at later times will be ahead (in the $u$ direction).
As a consequence of these two properties, we can define (for a given finite sample) the critical
force $f_c^{\tt{sample}}$ as the maximal force for which a metastable configuration still exists;
i.e., a configuration for which al points $x$ verify $\partial_t u(x,t)=0$.
Analogously to the ground state polymer for the statics (Sec.~\ref{sec:equilibrium}), this last 
{\it unique} metastable configuration displays self-affine properties with a well defined 
roughness exponent $\zeta_{dep}$ (see \fig{criticalstate}a).
For a finite sample, $f_c^{\tt{sample}}$ is a stochastic variable which 
depends on the disorder realization.
As the size $L$ of the sample grows, the asymptotic  critical threshold $f_c$  is approached
(see \fig{criticalstate}b), and the fluctuations of  $f_c^{\tt{sample}}$  are suppressed 
as~\cite{bolech_critical_force_distribution}
\begin{equation}
\overline{\left[f_c^{\tt{sample}}\right]^2}-\left[\overline{f_c^{\tt{sample}}}\right]^2  \sim L^{-2/\nudep} ;
\end{equation}
\noindent where $\nudep$ is the characteristic correlation length exponent.
Both the exponents $\zeta_{dep}$ and $\nu_{dep}$ can be carefully determined using numerical
algorithms that avoid the direct numerical integration of Eq.~(\ref{eq:eqmotion}).
The key idea behind these methods is to check for the existence of metastable states
for the string in the particular sampled disorder for a given value of the external force $f$.
This check can be performed in a computing time that grows only linearly with the system size
using properties discussed by 
Middleton~\cite{rosso_depinning_longrange_elasticity,rosso_correlator_RB_RF_depinning}.

\subsubsection{Critical and dynamical exponents: the non-steady relaxation}

The accurate determination of critical exponents is a difficult task, specially for critical 
phenomena in disordered systems.
The so-called Short-Time Dynamics method (STD) allows to obtain critical exponents values
without the need of system equilibration.
It relies on the validity of an homogeneous relation for the order parameter when the system
performs a non-steady relaxation at the critical point.
This scaling relation, in contrast with traditional equilibrium finite-size scaling methods,
can be used to avoid finite-size effects but includes dependences with time and initial 
conditions~\cite{JaScSc1989,Zh2006,review_albano}.

To apply the STD to the depinning transition we 
rely on the analogy with standard phase transitions~\cite{fisher_depinning_meanfield}.  
By considering the velocity $v$ as an order parameter, 
and the dimensionless force $(f-f_c)/f_c$ as the reduced driving field,
we expect the following homogeneity relation to be valid for the 
long-time relaxation of an initially flat string (infinite velocity initial 
condition\footnote{This is equivalent to the case of an ``ordered'' initial
condition for the order parameter.})~\cite{kolton_short_time_exponents,kolton_universal_aging_at_depinning,FeBuKo2013}:
\begin{equation}\label{eq:scaling2v}
    v(h,L,t) = t^{-\beta/\nu z} \tilde{v}_{\pm}(t^{1/z\nu} h,t^{-1/z} L) ;
\end{equation}
where $h = |f-f_c|/f_c$ and the function $\tilde{v}_{\pm}$ has two branches 
depending on the sign of $f-f_c$. 
Close to depinning, the relaxational dynamics described by Eq.~(\ref{eq:scaling2v}) 
is valid in a short-time regime, after which the velocity reaches a steady-state
value if $f>f_c$, while for $f<f_c$ the string is blocked in a metastable 
state (with memory of the initial condition) and $v \to 0$.
Exactly at the critical point ($h=0$) and in the limit $L\to \infty$ we 
expect a power-law behavior for the velocity, $v \sim t^{-\beta/\nu z}$. 
Alternatively, we can write $v \sim \ell(t)^{-\beta/\nu}$, where $\ell(t)\sim 
t^{1/z}$ is the dynamical length, describing the growth of roughness 
as $w(t) \sim \ell(t)^{\zeta} \sim t^{\zeta/z}$.
By fitting $v(t)$ and $w(t)$ we thus in principle get $\beta/\nu z$ and ${\zeta/z}$ 
independently. 
Since $\zeta$ (and therefore $\nu$) can be obtained independently using the method
described in Sec.~\ref{sec:criticalforce}, with the STD method it is possible to extract 
in addition the critical exponent $\beta$ and the dynamical exponent $z$.

In Ref.~\cite{FeBuKo2013} large scale Langevin dynamics simulations 
have been implemented to obtain $v(t)$ and $w(t)$ over a large period of time 
for large system sizes, such that $L \gg \ell(t)$. 
This has allowed to detect robust (under different shape
and nature of the disorder correlator) scaling corrections to the asymptotic 
power-law forms, and thus to avoid fitting 
effective power-law decays, yielding significantly biased incorrect exponents. 
A practical criterion to estimate the crossover between the ``mesoscopic'' time 
regime with corrections and the truly universal ``macroscopic'' time regime was 
found by observing that the hyperscaling relation $\beta=\nu(z-\zeta)$ is 
violated in the former regime. 
This can be easily seen by analyzing the quantity $\gamma(t) = \frac{w(t)}{t v(t)}$: 
if the measured $\gamma(t)$ depends on time, it means that the 
system has not reached the critical scaling regime yet. 

\begin{figure}[!tbp]
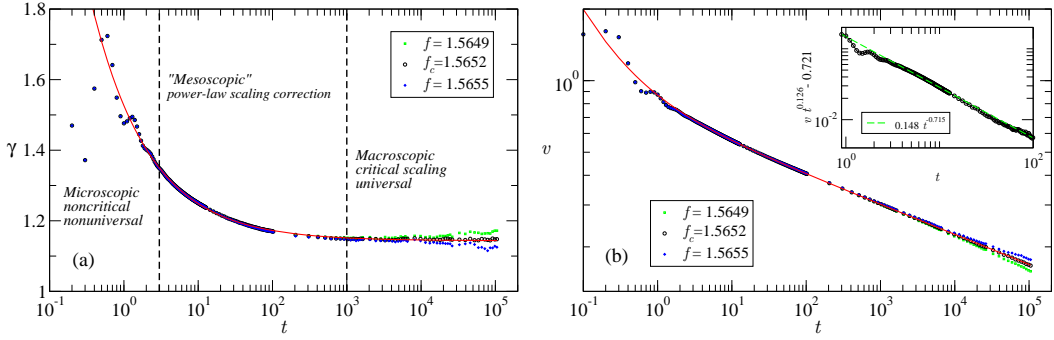

\centering
\includegraphics[scale=0.29,clip=true]{figures/gamma_vs_t}
\hfill
\includegraphics[scale=0.29,clip=true]{figures/v_vs_t}
\caption{\label{fig:L2e25RBgamma_and_v_vs_t_fitted}
 Evolution of $\gamma(t)$ (a) and $v(t)$ (b) for $f\approx f_c$, 
with $f_c$ the {\it thermodynamic} critical force. 
Two other forces, just above ($f_c^+$) and just below ($f_c^-$) $f_c$ are also shown. 
Continuous lines are fits for $\gamma(t)$ and $v(t)$ 
including power-law corrections to the asymptotic scaling yielding the true 
critical exponents. 
In (a) vertical dashed lines separate qualitatively different time regimes.
The inset in (b) shows that corrections to scaling are power-law like.
\textit{Figures are adapted with permission from}~\cite{FeBuKo2013}.
} 
\end{figure}

In \fig{L2e25RBgamma_and_v_vs_t_fitted}(a) we show the behavior of $\gamma(t)$. 
After a microscopic regime, $\gamma(t)$ slowly decreases implying an 
``effective'' unbalanced relation $\left[\nu(z-\zeta)-\beta \right]_{\mathrm{eff}}<0$.
In this mesoscopic regime the critical dynamics scaling is not valid. 
After a surprisingly long crossover however, $\gamma(t)$ develops a plateau. 
This information allows to determine the appropriate time regime of $v(t)$ and $w(t)$
where it is possible to fit $\beta/\nu z$ and $\zeta/z$, respectively.
In \fig{L2e25RBgamma_and_v_vs_t_fitted}(b) the behavior of $v(t)$ is shown.
Power-law corrections of the form
$v(t) = V_0 ~ t^{-\beta/\nu z} \left[ 1 + \left({t}/{t_v}\right)^{-\alpha_v} \right]$
(with $V_0$, $t_v$, and $\alpha_v$ fitting parameters) are find.
At large times, when $\gamma(t)$ develops a plateau, an accurate power-law fit can be 
performed for $v(t)$ at the thermodynamic depinning force $f_c=1.5652\pm0.0003$, yielding 
$\beta/(\nu z) = 0.128 \pm 0.003$. 
The observation of scaling corrections explain the systematically larger 
effective values for $(\beta/\nu z)$ reported before for smaller systems~\cite{FeBuKo2013}. 
An analogous analysis for $w(t)$ shows similar robust power-law scaling corrections, 
and gives a consistent value for the exponent, $1-\zeta/z=0.128\pm0.003$.
These results, combined with an independently determined value of $\zeta=1.250\pm0.005$
(see \fig{criticalstate}a) and the relation $\nu=1/(2-\zeta)$,
allows to get $\beta = 0.245 \pm 0.006$, $z = 1.433 \pm 0.007$, $\zeta=1.250 \pm 0.005$ 
and $\nu=1.333 \pm 0.007$~\cite{FeBuKo2013}. 
Scaling forms for the joint force-time dependence of the velocity extracted 
from Eq.~(\ref{eq:scaling2v}) are in excellent agreement with these exponents~\cite{FeBuKo2013}.

\subsection{Depinning regime at $T>0$: the critical thermal rounding}
\label{sec:thermal_rounding}

\begin{figure}[!tbp]
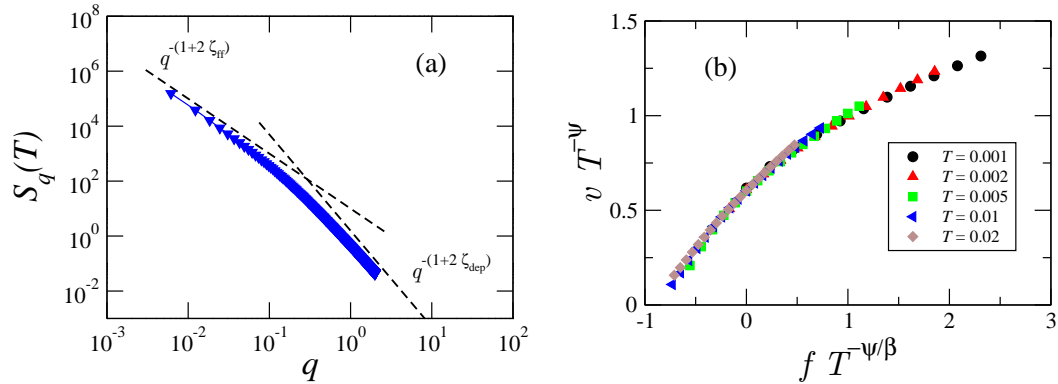

\centering
\includegraphics[scale=0.5,clip=true]{figures/Sq-temp}
\hfill
\includegraphics[scale=0.5,clip=true]{figures/vf-temp}
\caption{\label{fig:thermal_rounding}
(a) Structure factor $S_q$ at a finite temperature $T=0.02$ showing two regimes with different
roughness exponents, $\zetaff$ and $\zetadep$, above and below the crossover length scale $\lav$, respectively.
(b) Velocity-force scaling function in the thermal rounding region.
The disorder intensity is kept fixed and the velocity for different temperatures collapses in a single function.
\textit{Figures are adapted with permission from}~\cite{bustingorry_thermal_depinning_exponent}.
} 
\end{figure}

When the temperature is finite, there is no sharp transition between zero and
finite velocity regimes.
At forces around the critical value, $f \approx f_c$, a finite temperature value
smears out the transition, which is no longer abrupt. This thermal rounding of the
depinning transition can be characterized, exactly at the critical force $f=f_c$, 
by a power-law vanishing of the velocity with the temperature $v \sim T^\psi$,
where $\psi$ is the so-called the thermal rounding 
exponent~\cite{chen_marchetti,bustingorry_thermal_rounding_epl,bustingorry_thermal_depinning_exponent,middleton_CDW_thermal_exponent,nowak_thermal_rounding}.

At small length scales, $q\gg 1/\lav$, the structure factor shows the typical
roughness regime associated to depinning, i.e. $S_q \sim q^{-(1+2
\zetadep)}$, while at large length scales, $q \ll 1/\lav$,
geometrical properties are dictated by effective thermal fluctuations induced by the
disorder, i.e. $S_q \sim q^{-(1+2 \zetaff)}$, as shown in \fig{thermal_rounding}(a).
In the critical region the depinning correlation length is given by the velocity as
$\lav \sim v^{-\nu/\beta}$.
Thus, the depinning correlation length depends on the temperature only through the
velocity and in the thermal rounding regime~\cite{bustingorry_thermal_rounding_epl}
$\lav \sim T^{-\psi \nu/\beta}$.
With this information one can write for the structure factor
\begin{equation}
 \label{eq:scal-Sq}
 S_q = T^{-\psi \nudep (1+2 \zetadep)/\beta} s\left( qT^{-\psi
\nudep/\beta} \right),
\end{equation}
\noindent where the scaling function $s(y) \sim y^{-(1+2\zetaff)}$ for $y \ll 1$ 
and $s(y) \sim y^{-(1+2\zetadep)}$ for $y \gg 1$.
In Ref.~\cite{bustingorry_thermal_rounding_epl} it was shown that the structure factor
scales with the previous form.

All the information about the thermal rounding regime can be gathered in the expected
universal behavior of $v(h,T)$.
It is possible to show that assuming the homogeneous relation between the velocity and
both temperature and force, as it is usual for phase transitions, a universal function
follows.
If there were not strong finite size effects, in the vicinity of the critical region the
velocity should scale as
\begin{equation}
\label{eq:universalv}
 v T^{-\psi} \sim h_\pm \left( f T^{-\psi/\beta} \right),
\end{equation}
\noindent with $h_\pm(x)$ a universal function which can in principle be different above ($h_+$)
and below ($h_-$) threshold.
One expects that $h_+\sim y^\beta$ for $y\gg 1$.
Figure~\ref{fig:thermal_rounding}(b) shows this scaling relation for fixed disorder intensity and 
different temperatures as indicated.
This universal behavior has been numerically confirmed using Langevin dynamics simulations
of the QEW equation~\cite{bustingorry_thermal_depinning_exponent} and indirectly tested
in experiments of domain wall motion in 2D ferromagnets~\cite{Bustingorry_thermal_rounding_fitexp}.

\section{Conclusions and Perspectives}
\label{sec:conclusions}

The driven QEW model represents a minimal paradigmatic model for 
studying the dynamics of driven elastic systems in random media. 
In this paper we have reviewed a series of numerical methods 
specially developed for studying different and key properties of this system.
The results found in the literature, lead us to the physical picture discussed in 
Sec.~\ref{sec:transportandgeometry} and ilustrated in~\fig{whatcanwesay}.

The most challenging aspect of the problem remains the quantitative comparison with 
experiments. On one hand, it would be important to make quantitative predictions
for different elastic universality classes. 
Although we expect the same basic physics described here 
to emerge at large enough scales, it is important 
to accurately obtain the corresponding critical exponents and 
to understand the additional dynamical 
crossovers that may arise at intermediate length-scales.

At last, probably the least understood but still experimentally relevant phenomena
for the dynamics of interfaces in random media are: the occurrence of ``plasticity'' 
(displayed by overhangs and bubbles~\cite{repain_creep}), 
the effect of internal degrees of freedom 
(such as the ``spin phase'' coupled to the position of magnetic domain walls
~\cite{leekim_creep_corriente,lecomte_internal}), 
and the effect of ``structural relaxation''~\cite{jagla_kolton_earthquakes} 
in host materials. 
This may encourage the development of new models and novel efficient
numerical approaches in the field.

\section*{Acknowledgements}
We acknowledge very fruitful collaborations with T.~Giamarchi, 
P.~Le Doussal, W.~Krauth, G.~Schehr, K.~Wiese, E.~Albano, E.A.~Jagla, and P.~Paruch.
We also acknowledge discussions with A.~Fedorenko, L.~Cugliandolo, 
D.~Dominguez and J.~Ferre. 
We thank M.V.~Duprez for helping us with the schematic figures.

\bibliographystyle{elsarticle-num}
\bibliography{tfinita5,gpgpu,std}

\end{document}